\documentclass{article}

\usepackage{arxiv}

\usepackage[utf8]{inputenc} % allow utf-8 input
\usepackage[T1]{fontenc}    % use 8-bit T1 fonts
\usepackage{hyperref}       % hyperlinks
\usepackage{url}            % simple URL typesetting
\usepackage{booktabs}       % professional-quality tables
\usepackage{amsfonts}       % blackboard math symbols
\usepackage{nicefrac}       % compact symbols for 1/2, etc.
\usepackage{microtype}      % microtypography
\usepackage{lipsum}
\usepackage{graphicx}

\title{Effects of Assortativity on Consensus Formation with Heterogeneous Agents}

\author{
  Ece Çiğdem Mutlu, Ivan Garibay \\
    Complex Adaptive System Laboratory\\
  University of Central Florida\\
  Orlando, FL 32816 \\
  \texttt{ece.mutlu, igaribay@ucf.edu} \\
}

\begin{document}
\maketitle

\begin{abstract}
Understanding the consensus formation and exploring its dynamics play an imperative role in studies of multi-agent systems. Researchers are aware of the significant effects of network topology on the dynamical process of consensus formation; therefore, much more attention has been devoted to analyzing these dependencies on the network topology. For example, it is known that the degree correlation between nodes in a network (assortativity) is a moderator factor which may have serious effects on the dynamics, and ignoring its effects in information diffusion studies may produce misleading results. Despite the widespread use of Barabasi's scale-free networks and Erdos-Renyi networks of which degree correlation (assortativity) is neutral, numerous studies demonstrated that online social networks tend to show assortative mixing (positive degree correlation), while non-social networks show a disassortative mixing (negative degree correlation). First, we analyzed the variability in the assortativity coefficients of different groups of the same platform by using three different subreddits in Reddit. Our data analysis results showed that Reddit is disassortative, and assortativity coefficients of the aforementioned subreddits are computed as -0.0384, -0.0588 and -0.1107, respectively. Motivated by the variability in the results even in the same platform, we decided to investigate the sensitivity of dynamics of consensus formation to the assortativity of the network. We concluded that the system is more likely to reach a consensus when the network is disassortatively mixed or neutral; however, the likelihood of the consensus significantly decreases when the network is assortatively mixed.  Surprisingly, the time elapsed until all nodes fix their opinions is slightly lower when the network is neutral compared to either assortative or disassortative networks. These results are more pronounced when the thresholds of agents are more heterogeneously distributed. 
\end{abstract}

% keywords can be removed

\section{Introduction}
Networks are representations of the connection patterns of complex systems in which entities might be proteins, individuals, economic goods, etc. Topological structures and different properties of diverse complex networks have been investigated for decades. Most importantly, the characterization of the mixing patterns of these network structures helps us to understand the evolutionary, functional and dynamic process of those complex systems \cite{allen2017two}. If nodes in a network tend to associate with other similar nodes, this pattern is called assortative mixing (also known as \textit{homophily}). The concept of assortativity is extensively studied since its introduction by Newman \cite{newman2002assortative} in 2002. Although its application areas are diverse, the assortativity of a network is generally determined by the Pearson correlation coefficient between the degree distribution of its nodes. Previous studies show that non-social networks generally show a disassortatively-mixed pattern, i.e. metabolic pathways, protein-protein interactions, power-grid, World-Wide-Web \cite{newman2003mixing} or yeast genes and proteins \cite{lee2015modes}; however, brain connections are assortatively mixed \cite{im2014altered} despite its non-social property. Social networks, on the other hand, tend to be assortative, i.e. Facebook \cite{ugander2011anatomy}, Flickr, mySpace \cite{hu2009disassortative}; however, there are some exceptions such as disassortative mixing patterns on Twitter \cite{wang2014whispers} and Youtube \cite{hu2009disassortative}. In such platforms where social networks are established, it is not surprising that people interact with other people similar to them. This similarity might be based on age, race, language, education or number of connections established. Fisher et al. argued that social networks are assortative only when they are built as a group-based network \cite{fisher2017perceived}. To either test the validity of this assumption and to understand the variability in the assortativity of social networks better, we evaluated the mixing patterns of three different subreddits and observed that assortativity values vary even in the different groups of the same platform. Although measuring assortativity will not give an idea about the variation of all entities in a network, it is useful in understanding the average mixing behavior of them; thus, it plays an important role in understanding the dynamics of epidemic spreading, signal connections, information diffusion or consensus formation in a system.

In this paper, we will be focusing on the effect of the average mixing behavior of heterogeneous agents on consensus formation. It is important to obtain a better understanding of consensus formation, since beliefs and opinions in social groups, including the society, constantly evolve as societal dynamics introduce paradigmatic shifts over time. Fashion trends, cultural changes, the rise and fall of political ideologies, marketing practices and technology innovations are good examples for these paradigm shifts leading to consensus formation. Simultaneous to these external influences in the society; internal communication patterns, such as online social networking activities of individuals, can also influence opinion formation, adoption, and dissemination of agents. For instance, it would be unexpected for an individual to adopt an opposing opinion in a network that predominantly supports another opinion. These internal communication patterns strongly depend on the network topology. Consensus formation of multi-agent systems agents have attracted researchers from many different disciplines. This concept is applied in many areas from spacecraft \cite{vandyke2006decentralized} to robotic teams \cite{arkin1997cooperative}. There are applications of consensus strategies in decision-making, the polarization of people in the examples of political affiliation \cite{jin2017political} and rumor spreading \cite{borge2012absence}. The multiplicity and diversity of uses necessitate a better understanding of the consensus formation process.

\section{Assortativity in Reddit}
As aforementioned, Twitter and Youtube show disassortative mixing patterns while Facebook, Flickr, and mySpace are assortative. This variability brings the following question that ``Do the assortativity values of the different groups of the same social network vary?''. The best example of the existence of different groups in a single platform is Reddit, in which authors can post their texts, URLs, images or videos, and/or add comments to other author's contents. Since contents are organized according to different subjects by user-created boards, which are called ``subreddit'' and thus, members can follow threads only if they are interested in it. Therefore, we collected all user activities in three different subreddits.

\begin{figure}[htbp]
\centering
\includegraphics[width=0.8\linewidth]{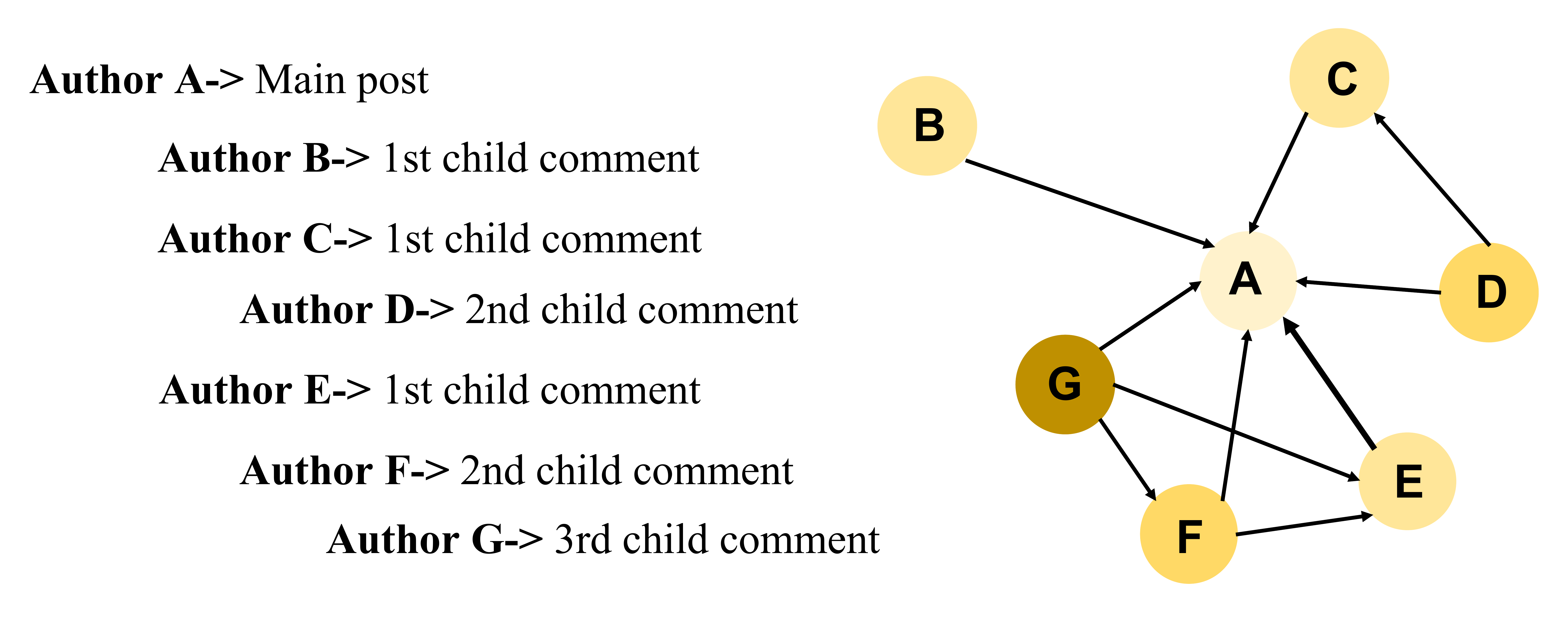}
\caption{A representation of Reddit network generation from a cascade.}
\end{figure}

Since there is no following or friendship in Reddit, we generated the Reddit network from the conversation patterns of users in subreddits. Suppose that we have a cascade in which one author posted (\textit{Author A}) and 6 other authors added a comment (\textit{Author B-G}) as shown in Figure 1. Since commenter authors reply \textit{Author A}'s post, we assume that \textit{Author A} is followed by all other users in the network (same as Twitter, if you follow other users, you can see their tweets). Since \textit{Author D} adds a comment to \textit{Author C}'s comment but not \textit{Author B}'s comment, we assume that \textit{Author B} has no influence on \textit{Author D}, and \textit{Author D} follows \textit{Author C} but does not follow \textit{Author B}. In the last three comments of the cascade, on the other hand, \textit{Author G} follows both \textit{Author E} and \textit{Author F} since cascade flows from parent to children comments continuously.

\begin{figure}[htbp]
\includegraphics[width=0.9\linewidth]{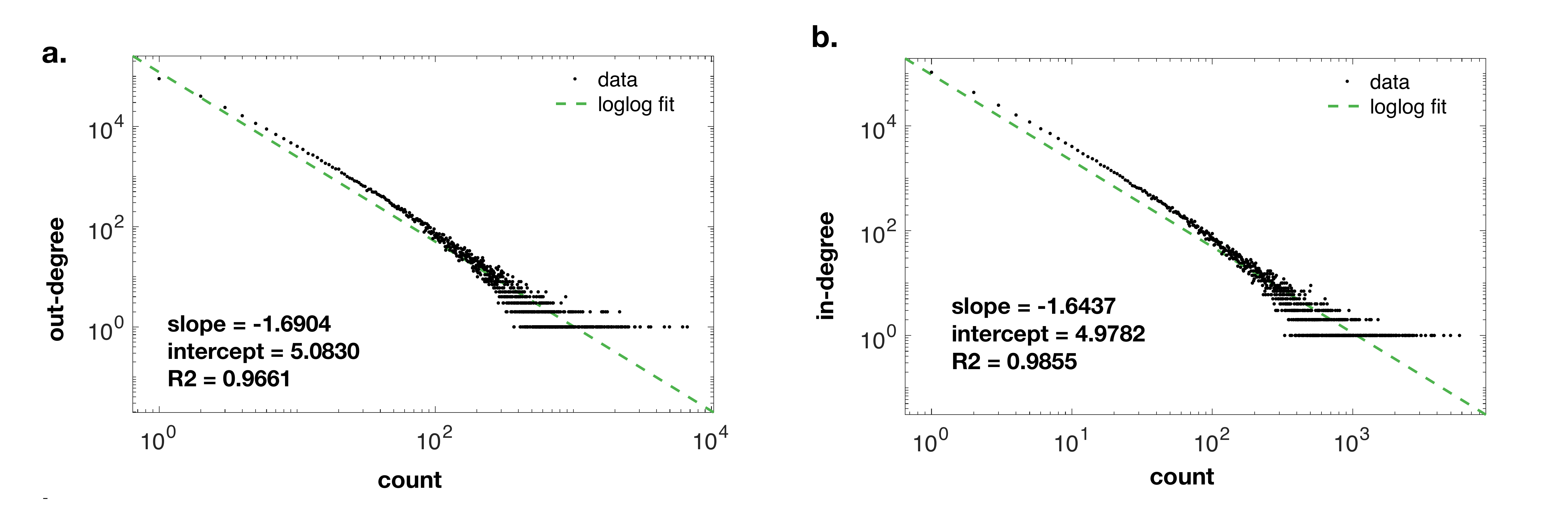}
\caption{Power-law fitness of a. out-degree b. in-degree of Reddit network from the most-populated subreddit.}
\end{figure}

We generated three different networks from three cybersecurity-related subreddits from January 2015 to September 2017. Since we created our network based on an assumption, we first checked the power-law fitness of in-degree and out-degree distribution of the network. Many studies demonstrated that the majority of the social networks are scale-free networks \cite{aparicio2015model}. High $R^2$ values show that in-degree and out-degree of the generated Reddit network show a great power-law fitness (Figure 2), which supports the validity of our assumption. After generating networks, we calculated assortativity coefficients of these subreddits (Table 1). Motivated by the variability in the values even in the same platform, we decided to investigate the sensitivity of dynamics of consensus formation to the assortativity of the network. 

\begin{table}[htbp]
  \caption{Details of Reddit datasets from 3 Subreddits}
  \label{tab:data}
  \centering
  \begin{tabular}{|l |c |c| c|}
    \hline
     & Subreddit1 & Subreddit2 & Subreddit3\\
    \hline
    Number of user-generated contents &16,332,268&1,446,083 &162,506\\
    \hline
    Number of unique users (\textit{Author})&751,561&119,088&33,762\\
    \hline
    Assortativity coefficient ($r$)&-0.0384&-0.0588&-0.1107\\
    \hline
\end{tabular}
\end{table}

\section{Method}
\subsection{Information Diffusion Models}
There are two main models used for information diffusion, which are Independent Cascade Model (ICM) \cite{goldenberg2001talk} and Linear Threshold Model (LTM) \cite{granovetter1983threshold}. Suppose that $G \langle V, E \rangle $ is a complex network (graph), which is defined as the set of vertices (nodes) ($V = \{v_1, v_2,...,v_n\ | n \in N* \}$) and edges between them ($V_{ij} = (v_i, v_j)$ where $(i,j \in N*; i \neq j, \forall i, \forall j <n)$). In ICM, individuals are assumed as bounded-rational, and their adoption can be predicted from the influence on each other. Here, each edge ($V_{ij}$) has its own influence probability ($p_{ij}$) which are previously known, and each node ($i$) has a single chance to activate its neighbor $j$ with probability $p_{ij}$. As in the case of LTM, individuals adopt a new opinion only if a critical fraction of their neighbors have already adopted the new opinion; thus, every node ($i$) is associated with a threshold value ($\phi_i$) to be activated in the next step. 

Despite the necessity of assigning heterogeneity to thresholds in LTM, this heterogeneity is poorly-defined among researchers, which leads to extensive use of uniform \cite{liu2018impacts}, \cite{sprague2017evidence}, \cite{singh2013threshold} and binary \cite{wang2016dynamics} thresholds in many studies. Arguably, this assumption of homogeneous or binary thresholds is an oversimplification of reality and may produce misleading results. To remedy this oversimplification and thereby provide more holistic and possibly more accurate models, many studies employed more complex threshold functions such as the tent-like function \cite{zhu2018dynamics}, the truncated normal distribution function \cite{karampourniotis2015impact} or the sigmoid function \cite{fink2016investigating}. In the current study, we employ LTM to understand the dynamics of opinion formation and control the threshold heterogeneity with a parameter $N_{th}$ during the simulations. 

\subsection{Simulation Procedure}
Although the main analysis here is to understand the effect of the assortativity of the network on the dynamics of consensus formation; we also investigated the effect of threshold heterogeneity to justify the robustness and scalability of the results. The heterogeneity of the agents is yielded by the diversity of their thresholds.

We first generated a random network using a configuration model, in which out-degree distribution is assigned as random numbers drawn from power-law distribution in the form of $\sqrt{N}x^{\gamma}$. Here, $\gamma=3$ and $N$ denotes the number of nodes in a network, that equals to $1000$. Furthermore, we keep the in-degree as constant ($k_{in}=17$), motivated by Dunbar number, i.e. individuals have a cognitive limit on the number of their social relationships \cite{dunbar1992neocortex}. This random network tends to be neutral (uncorrelated) or slightly disassortative. To tune the magnitude of the assortativity, we applied the Xulvi Brunet-Sokolov rewiring algorithm \cite{xulvi2005changing}. This algorithm chooses two linked node pairs at each time step, i.e. $i,j$ and $m,n$ where $A_{ij}, A_{mn}=1$ and $A$ denotes the adjacency matrix. Then, it orders these four nodes according to their degree, i.e. Suppose that $k_i<k_m<k_j<k_n$. To increase the assortativity, first two nodes and last two nodes i.e. $A_{ij}, A_{mn}=0$  and  $A_{im}, A_{jn}=1$; to decrease the assortativity, first node with the last node and second node with the third node are rewired by destroying the previous linkage, i.e. $A_{ij}, A_{mn}=0$  and  $A_{in}, A_{mj}=1$. This process continues until the desired assortativity is obtained. Note that, this algorithm does not change the overall degree distribution, thus mean a degree in the network; however, a rewired network may exhibit different geometrical and transport properties. The Xulvi Brunet-Sokolov algorithm considers only the out-degree distribution of the directed graph since the in-degree is constant.

In the next step, we initialized the opinions of individuals as a Bernoulli distributed random variable with an initial probability ($p$), i.e. the opinion of the node $i$ ($s_i$) might equal to $1$ with a probability $p$ and equal to $0$ with probability ($1-p$). This probability value of $p$ has a range of $0.2$ to $0.8$ in the current study. 

The heterogeneity of the agents is yielded by the heterogeneity in their thresholds of adopting a new opinion. For this purpose, we assigned thresholds as a uniformly distributed random variable ($\phi_i \in unif(0.5, 1)$) throughout the interval defined by $N_{th}$ which takes value from the set of ${2,5,10,100}$ and increasing $N_{th}$ yield more heterogeneity among agents. Thresholds of the agents are randomly assigned from the subset defined by:

\begin{equation}
    \phi_i \in \{0.5, 0.5 + \frac{0.5}{N_{th}-1}, ..., 0.5 + \frac{0.5 (N_{th}-2)}{N_{th}-1},1\}
\end{equation}

After generating the network and bringing its assortativity to the desired degree, initializing the opinions and assigning thresholds, we run the opinion change simulations. The process of updating opinions is as follows as in \cite{mutlu2020degree}:

\begin{enumerate}
    \item Picking a node $i$ randomly.
    \item Calculating the weighted average of the opinions of its in-neighbors ($\bar{o}_i$). Here, weights are the multiple edges formed between node $i$ and its neighbors. 
    \item Updating the opinion of node $i$ (${s}_i$) according to the criteria as follows:
    \begin{enumerate}
        \item \textbf{if} ${s}_i=0$ and $\bar{o}_i-{s}_i > \phi_i$, 
        \newline 
        \textbf{then} ${s}_i=1$ in the next step.
         \item \textbf{if} ${s}_i=1$ and $\bar{o}_i-{s}_i<-\phi_i$, 
         \newline 
         \textbf{then} ${s}_i=0$ in the next step.
    \end{enumerate}
\end{enumerate}

This Markovian chain is repeated until individuals fix their opinion. 

\section{Results}
In the current study, we first aimed to analyze the effect of change in the network mixing pattern on the average opinion at steady state ($\bar{s}$). Therefore, after all the individuals fix their opinions in the network, we averaged their opinions by using the equation below:
\begin{equation}
    \bar{s}=\frac{1}{N}\sum_{i}^{N}{s_i(\infty)}
\end{equation}

where $s_i(\infty)$ is the opinion of node $i$ at steady-state and $N$ is the seed size, i.e. number of nodes in the network. We conducted our simulations to measure $\bar{s}$ as a function of the initial probability ($p$) with a varying assortativity ($r$) and varying threshold heterogeneity $N_{th}$.

\begin{figure}[htbp]
\includegraphics[width=\linewidth]{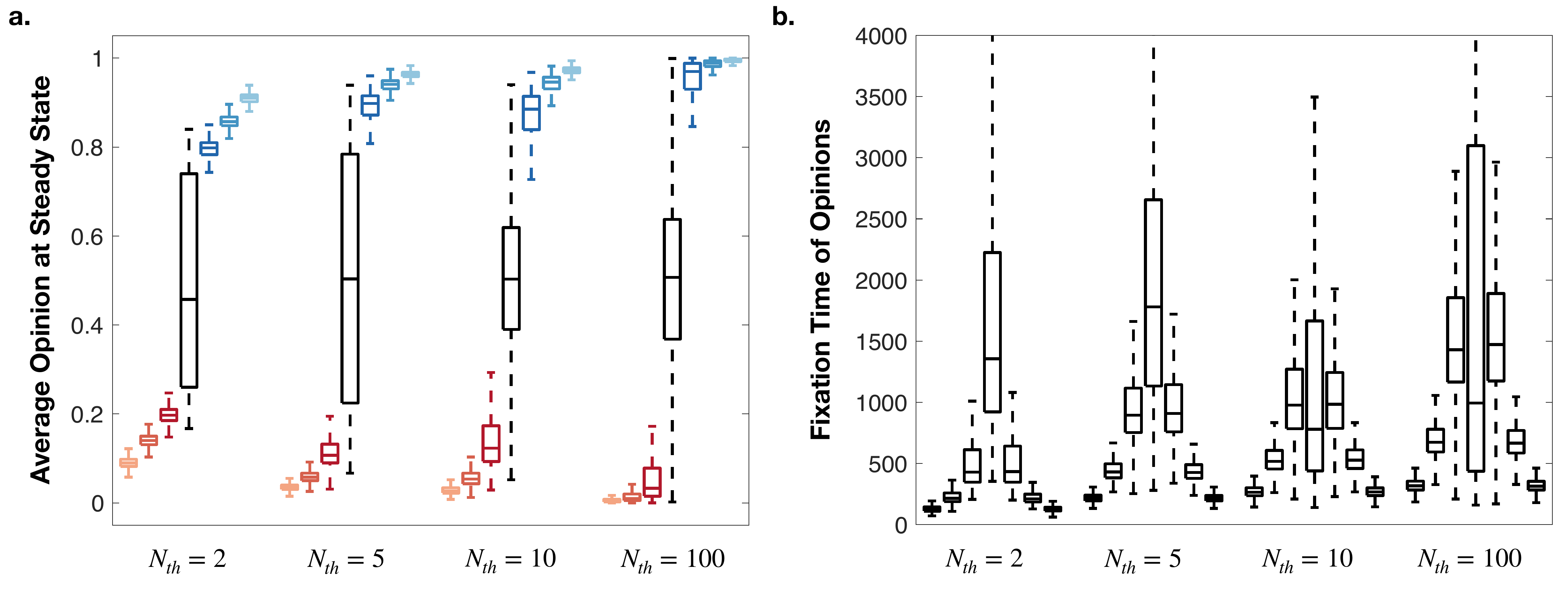}
\caption{a. Average opinion at steady state ($\bar{s}$) b. Fixation time of opinions ($t_F$) when $N=1000$, $r=0$ and $N_{th}=\{2,5,10,100\}$. For each $N_{th}$ seven boxplots show the variation in $p={0.2,0.3,...,0.8}$.} 
\end{figure}

Box plots in Figure 3 show the distribution of repeated experiments on the average opinion at the steady-state ($\bar{s}$) (left) and the time elapsed until all individuals fix their opinion ($t_F$) (right) when $N=1000$. We fixed assortativity coefficient of the network as neutral ($r=0$) since many studies in the literature generates Barabasi's scale-free and/or Erdos-Renyi networks, which are tend to show uncorrelated mixing pattern. Each figure includes 4 groups of box plots, in which groups represent various $N_{th}$, and $p$ varies from 0.2 to 0.8 in each group. Since average opinion shows the sample mean of the opinions at steady-state, the values closer to either 0 or 1 show the dominance of one of the opinions, i.e. the system is more likely to reach a consensus. Figure 3a shows that $p$ determines the dominance of the opinions; when $p \leq 0.4$, $\bar{s}$ take values close to 0, while $p \geq 0.6$ it approaches to 1. Furthermore, there is a clear asymmetry before and after $p=0.5$ in all cases. As $p$ closes to 0.5, system has mix of both opinions rather than reaching a consensus. Figure 3b, on the other hand, shows the relatively long duration of opinion change process when opinions are initialized halfway ($p = 0.5$). Although threshold heterogeneity of nodes in the system has a slight effect in the resulting average opinion when thresholds are out-degree dependent, we can conclude that the probability that the system reaches a consensus slightly increases as the threshold heterogeneity increases; however, the time elapsed until all nodes fix their opinions increases significantly with increasing $N_{th}$.

Figure 4, on the other hand, shows the effect of rewiring the network before opinion update simulations, when network is assortative ($r=0.1$), neutral ($r=0$) and disassortative ($r=-0.1$). All simulations are carried out at $N_{th}=5$ and $N_{th}=10$ for each mixing pattern to understand the moderator effect of the threshold heterogeneity.
Results show that the increase in the assortativity coefficient of a network has a prominent effect on the average opinion at the steady state at both threshold heterogeneity. The system is more likely to reach a consensus when network is disassortatively mixed and this effect is more prominent when thresholds are more heterogeneous. Surprisingly, $t_F$ reaches its minimum value when network is neutral. Figure 4b shows that bringing the network to the steady-state takes more time when degree distribution of its nodes either positively or negatively correlated. Again, this unexpected effect is more prominent when threshold heterogeneity is high. 

\begin{figure*}[htbp]
\includegraphics[width=\linewidth]{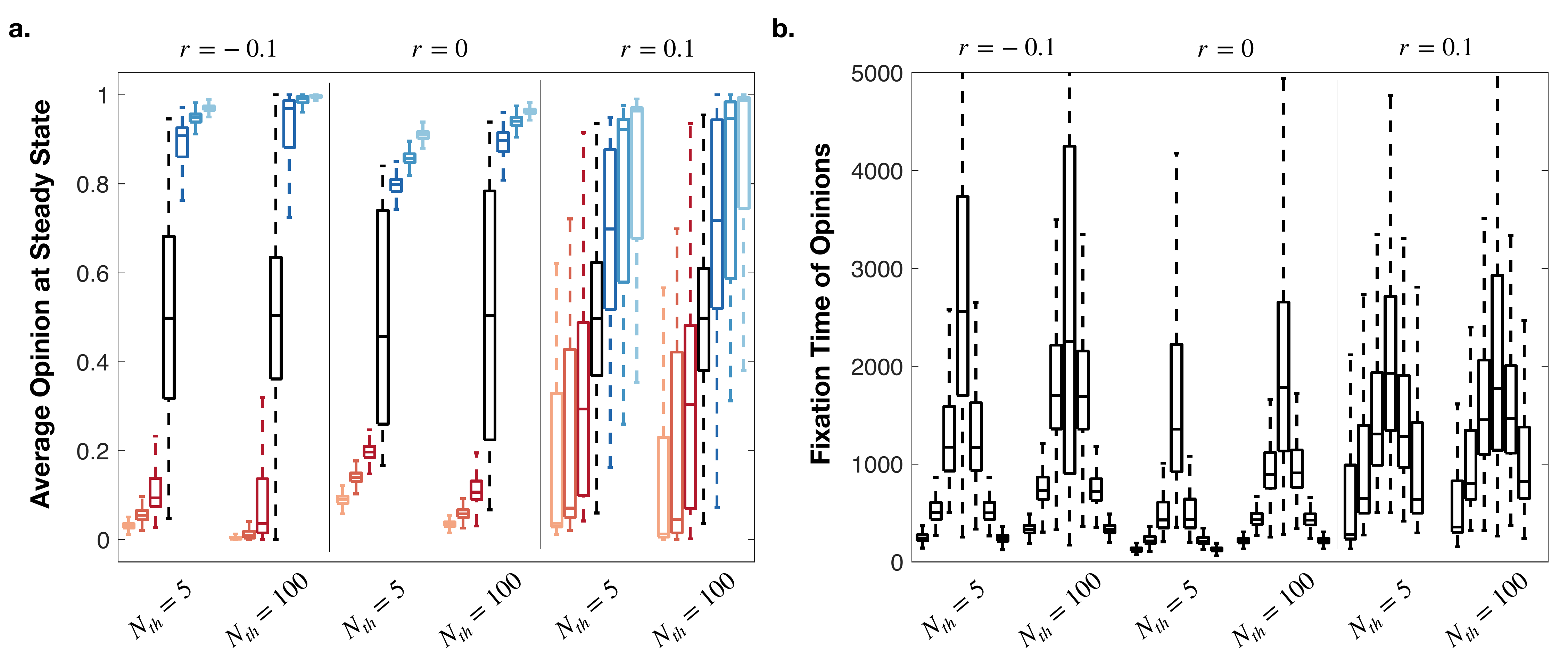}
\caption{a. Average opinion at steady state ($\bar{s}$) b. Fixation time of opinions ($t_F$) when $N=1000$, $r=-0.1$ in dissortative, $r=0$ in neutral and $r=0.1$ in assortative mixing. For each $N_{th=5}$ or $N_{th=100}$, and seven boxplots show the variation in $p={0.2,0.3,...,0.8}$.} 
\end{figure*}

\begin{figure*}[ht]
\includegraphics[width=\linewidth]{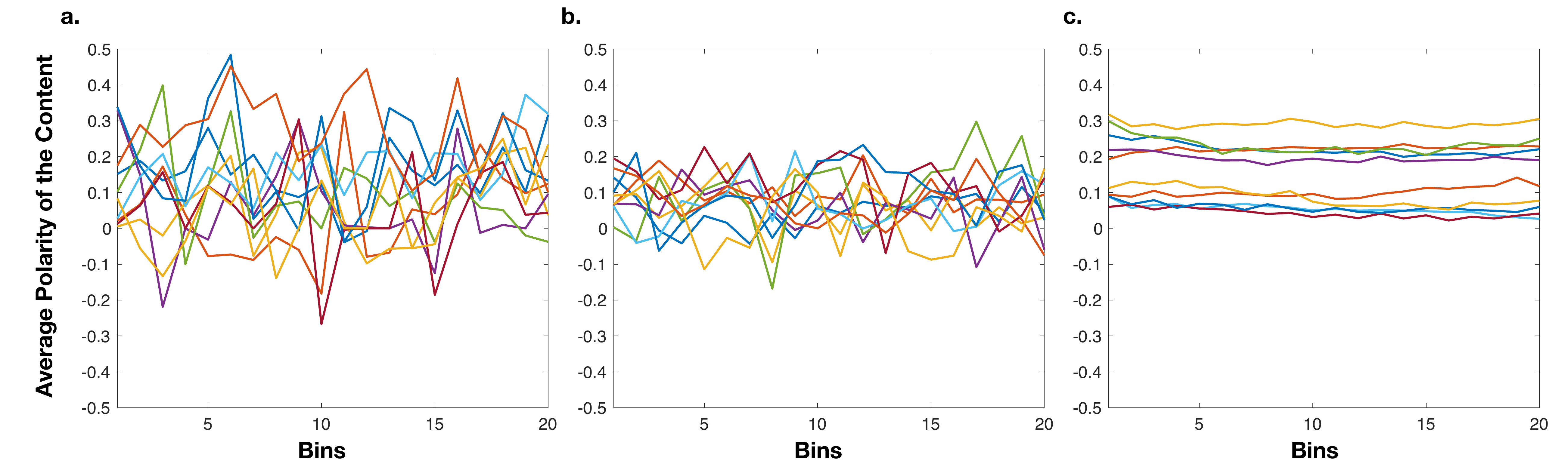}
\caption{Average polarization of the user-generated contents from 10 largest cascades in a. Subreddit1, b. Subreddit2 c. Subreddit3} 
\end{figure*}

To validate our results, we observed the opinion change of users in 10 largest cascade of three aforementioned subreddits. We considered the sentiment polarity ,which is a float within a range [-1.0,1.0], for every user-generated content by using CLIPS's sentiment function \cite{clips_2010} as a opinion towards the content of the discussion. Since each cascade have different size, we grouped (i.e. by averaging) the polarity scores into 20 bins after sorting them according to their time. We investigated the fluctuations in the polarity scores for each cascade in Subreddit1 ($r=-0.0384$), Subreddit2 ($r=-0.0588$) and Subreddit3 ($r=-0.1107$). Figure 5 shows the fluctuations in the average opinion at every time step in these subreddits. When network is less disassortative (close to neutral), opinion fluctuates excessively (Figure 5a); while opinions show a relatively stable pattern when network is more disassortative (Figure 5c). Since thresholds of users in Reddit are ambiguous but we concluded that threshold heterogeneity has a slight effect on reaching the consensus; we can argue that these real-world results match with our simulation results.

\section{Conclusion}
It is important to understand the dynamics of consensus formation in multi-agent system studies. Opinion formation and change depends on either the external effects, i.e. change in belief, ideologies and/or technology, or internal effects, i.e. change in interactions due to evolving network structure, thresholds to adopt a new opinion. The mixing pattern of the networks has been studied for many years, however, its effect on the dynamics of consensus formation is not analyzed in detail. In the current study, we investigated the sensitivity of dynamics of consensus formation to the assortativity of the network in the existence of heterogeneous agents. The contribution of the paper is three-fold: First, we created a Reddit network based on the conversation pattern of the users and validated our assumption with the power-law fitness of degree distributions. Our results show that Reddit topology can be created by using the post-comment relationships. Second, we explored the variability in the assortativity of sub-networks in the same platform. Although assortativity coefficients of many networks are available in the literature, none of the studies give information about Reddit network to our knowledge. For this purpose, we calculated the assortativity coefficients of the three subreddits and found that those values vary greatly, but all show disassortatively mixed behavior. Third, we examined the effect of assortative mixing in networks on the dynamics of consensus formation with multi agent-based simulations. During the simulations we tested the effect of assortativity coefficient of network, initial probabilities of the different opinions and threshold heterogeneity of the agents in the network structure. We concluded that the system is more likely to reach a consensus when the network is disassortatively mixed or neutral; however, the likelihood of the consensus significantly decreases when the network is assortatively mixed.  Surprisingly, the time elapsed until all nodes fix their opinions is slightly lower when the network is neutral compared to either assortative or disassortative networks. Reaching the consensus is more likely but more time-consuming when thresholds of agents in the system are more heterogeneous. This slight effect of heterogeneity is observed every cases regardless of the mixing pattern of the nodes; however, its positive effect reaching a consensus is more pronounced when the system is disassortatively mixed. We also validated our findings with real world Reddit data. 

\section{Acknowledgment}
This work is partially supported by grant FA8650-18-C-7823 from the Defense Advanced Research Projects Agency (DARPA).

\bibliographystyle{unsrt}  
\bibliography{references}  %%% Remove comment to use the external .bib file (using bibtex).
\end{document}